\documentclass[aps,prl,twocolumn,superscriptaddress,showpacs,amsmath,amssymb]{revtex4-1}%
\usepackage{amssymb}
\usepackage{amsfonts}
\usepackage{amsmath}
\usepackage{graphicx}%

\begin{document}
\title{Magnetic field - dependent Labusch parameter in LiFeAs superconductor from Campbell penetration depth}
\author{Plengchart~Prommapan}
\affiliation{The Ames Laboratory, Ames, IA 50011}
\affiliation{Department of Physics \& Astronomy, Iowa State University, Ames, IA 50011}
\author{Makariy~A.~Tanatar}
\affiliation{The Ames Laboratory, Ames, IA 50011}
\author{Bumsung Lee}
\affiliation{CeNSCMR, Department of Physics and Astronomy, Seoul National University, Seoul
151-747, Republic of Korea}
\author{Seunghyun Khim}
\affiliation{CeNSCMR, Department of Physics and Astronomy, Seoul National University, Seoul
151-747, Republic of Korea}
\author{Kee Hoon Kim}
\affiliation{CeNSCMR, Department of Physics and Astronomy, Seoul National University, Seoul
151-747, Republic of Korea}
\author{Ruslan~Prozorov}
\email{Corresponding author: prozorov@ameslab.gov}
\affiliation{The Ames Laboratory, Ames, IA 50011}
\affiliation{Department of Physics \& Astronomy, Iowa State University, Ames, IA 50011}

\date{7 March 2011}

\begin{abstract}
A \textquotedblleft true \textquotedblright\ critical current density, $j_{c}$, as opposite to
commonly measured relaxed persistent (Bean) current, $j_{B}$, was extracted from the Campbell penetration depth, $\lambda_{C}(T,H)$ measured in single crystals of LiFeAs. The effective pinning potential is
non-parabolic, which follows from the magnetic field - dependent Labusch parameter $\alpha$. At the equilibrium (upon field - cooling), $\alpha\left(  H\right)$ is non-monotonic, but it is monotonic at a finite gradient of the vortex density. This behavior leads to a faster magnetic relaxation at the lower fields and provides a natural
\emph{dynamic} explanation for the fishtail (second peak) effect. We also find the evidence for strong pinning at the lower fields. The inferred field dependence of the pinning potential is consistent with the evolution from strong pinning, through collective pinning and, eventually, to a disordered vortex lattice. The values of $j_{c}\left(  2\text{ K}\right)  \simeq 2\times10^{6}$ A/cm$^{2}$ provide an upper estimate of the current carrying capability of LiFeAs. Overall, vortex behavior of almost isotropic, fully-gapped LiFeAs is very similar to highly anisotropic d-wave cuprate superconductors, the similarity that requires further studies in order to understand unconventional superconductivity in cuprates and pnictides.

\end{abstract}

\pacs{74.25.Ha,74.25.Op,74.70.Xa,74.70.Ad}
\maketitle


The determination of the critical current density $j_{c}$, is one of the fundamental problems in the vortex physics of type-II superconductors. Not only it is important for the assessment
of the current-carrying capabilities relevant for practical applications, but
knowing \textquotedblleft true\textquotedblright\ $j_{c}$ is needed to
understand microscopic mechanisms of vortex pinning. What is often called
\textquotedblleft critical current\textquotedblright\ is routinely determined
from conventional DC magnetization measurements, alas this quantity is a
convolution of \textquotedblleft true\textquotedblright\ $j_{c}$ and magnetic
relaxation during the characteristic time, $\Delta t$, of the experiment. For
example, in case of ubiquitous \textit{Quantum Design} MPMS (SQUID)
magnetometery, $\Delta t\geq10$ sec. We will call measured supercurrent $j_{B}$ to distinguish it from the \textquotedblleft true\textquotedblright\ $j_{c}$ that is achieved when the vortices are de-pinned by the Lorentz force. By
definition, $j_{c}$ is reached when the energy barrier for vortex motion
vanishes, $U\left(  j_{c}\right)  =0$, whereas the measured current density
$j_{B}$ is determined by $U\left(  j_{B}\right)  =k_{B}T\ln\left(  1+\Delta
t/t_{0}\right)  $, where $t_{0}\lesssim1$ $\mu$sec is the characteristic time
scale that depends on both sample geometry and details of pinning
\cite{Geshkenbein1989,Vinokur1991,Blatter1994,Yeshurun1996,Burlachkov1998}.
This also results in a quite different temperature dependence of $j_{B}\left(
T\right)  $ compared to $j_{c}\left(  T\right)  $. Another approach to measure critical current density is to use AC susceptibility. Conventional time-domain susceptometers operate at frequencies $f \lesssim 10$
kHz (hence $\Delta t\gtrsim0.1$ msec) and have large driving amplitudes,
$H_{ac} \gtrsim 0.1$ Oe. Such perturbation displaces vortices from the potential
wells and one can use harmonics analysis to determine frequency - dependent
current density, $j_{B}\left(  T,B,f\right)$. This technique has been
applied in both global \cite{Shatz1993} and local
\cite{Prozorov1994a,Prozorov1995} forms.

In Fe-based superconductors
flux creep is substantial at all temperatures, thus measured $j_{B}$ is expected
to be lower than $j_{c}$. Indeed, reports produce only moderate current
densities, $j_{B}\lesssim10^{6}$ A/cm$^{2}$, - unusual for low-anisotropy
high$-T_{c}$ materilas
\cite{Prozorov2008,Yang2008,Yang2008a,Prozorov2009,Kim2009BaK122vortex,Shen2010,Pramanik2010LiFeAs}.

To access the information about pinning potential itself, one needs to measure
the linear response when vortices are not driven out of the pinning potential
wells. One way to do this is to measure so-called Campbell penetration depth
which determines how far a small AC magnetic field penetrates the
superconductor in the presence of vortices (induced by static external
magnetic field) in the limit of $H_{ac}\rightarrow0$, when vortex response is
purely elastic and linear \cite{Brandt1991,Brandt1995,Koshelev1991}. For a
pinning potential, $V\left(  r\right)$, the vortex displacement from the equilibrium position due to small $H_{ac}$ is found from $dV/dr=f_{L}$, where the Lorentz force, $f_{L}=j\times\phi_{0}/c$.Maximum force determines the
\textquotedblleft true\textquotedblright\ critical current density,
$j_{c}=c\alpha r_{p}/B$, attained at the range of the pinning potential
$r_{p}$. If vortex distribution is inhomogeneous, static (Bean) current
\cite{Bean1964}, $j_{B}$, is superimposed with the excitation AC current and
the response is determined by the effective Labusch constant $\alpha\left(  j_{B}\right) \equiv  =\left.
d^{2}V/dr^{2}\right\vert _{r=r_{0}}$. Clearly
$\alpha\left(  j_{B}\right)  $ is constant only for a parabolic $V\left(
r\right)  $. The Campbell penetration depth is given by $\lambda_{C}^{2}%
=\phi_{0}B/\left(  4\pi\alpha(j_{B})\right)$
\cite{Brandt1991,Brandt1995,Koshelev1991,Prozorov2003}.

Consider a typical experiment, which we use in the following. Sample is cooled in zero magnetic field and then static magnetic field is applied. This creates a gradient of vortex density supported by the
persistent Bean current density $j_{B}$ \cite{Bean1964}. Small-amplitude
$H_{ac}$ causes vortex vibrations within pinning potential well, a condition for Campbell
penetration depth measurements \cite{Brandt1991,Brandt1995,Koshelev1991,Prozorov2003}. After the sample is
warmed above $T_{c}$ it is cooled again keeping external static field constant
(field-cooling) whence $j_{B}=0$. We therefore may expect some hysteresis with
$\lambda_{C,ZFC}>\lambda_{C,FC}$ if $V\left(  r\right)  $ is non-parabolic.
Therefore, by measuring zero field - cooled (zfc) field-cooled (fc)
$\lambda_{C}$ at different magnetic fields and temperatures we can estimate
\textquotedblleft true\textquotedblright\ $j_{c}\left(  H,T\right)  $ and
access the information regarding shape of the pinning potential. For more
details the reader is referred to earlier studies of high $-T_{c}$ cuprates
\cite{Prozorov2003}.

One of the most interesting and commonly observed features of unconventional
superconductors is so-called second magnetization peak (also known as
\textquotedblleft fishtail\textquotedblright) \cite{Blatter1994}. It has now
been observed in most Fe-based superconductors when magnetic field is aligned
parallel to the crystallographic $c-$ axis
\cite{Prozorov2008,Yang2008,Prozorov2009,Kim2009BaK122vortex,Beek2010PrNd1111,Shen2010,Pramanik2010LiFeAs}. The origin of fishtail can be static, i.e., when \textquotedblleft
true\textquotedblright\ $j_{c}\left(  H\right)  $ is a non-monotonic function
of field, $H$, or it can be dynamic caused by field-dependent magnetic relaxation \cite{Burlachkov1998,Mikitik2001}.
Experimental determination of the origin of the fishtail in each material is,
thus, very important as it allows to shed light on the nature of the flux
pinning, hence defect structure \textquotedblleft seen\textquotedblright\ by
the Abrikosov vortices. In Fe-based superconductors, the interest is further
fueled by multiple reports that defects, even non-magnetic, are pair-breaking
due to, presumably, unconventional $s_{\pm}$ symmetry of the order parameter
\cite{Kogan2009,Gordon2010}. Additionally, it seems that low-field behavior of
most pnictides is governed by the so-called strong pinning, which results in a
sharp peak in magnetization at $H\rightarrow0$ \cite{Beek2010PrNd1111}.
Therefore, to conduct a clean, baseline experiment, one ideally needs Fe-based
superconductor with reduced scattering. These materials are rare, but do
exist in form of only few stoichiometric compounds, LiFeAs being one of them.
Due to high sensitivity to air and moisture, there are only few
reports on the vortex properties in LiFeAs crystals. Fishtail effect and
relatively high $j_{B}\left(  5\text{ K}\right)  \approx1\times10^{5}$
A/cm$^{2}$ were found in Ref.\cite{Pramanik2010LiFeAs}, whereas much lower
$j_{B}\left(  5\text{ K}\right)  \approx 1\times10^{3}$ A/cm$^{2}$ was reported
in Ref.\cite{Song2010}. Such spread may be related to clean - limit
superconductivity in this compound when even small variation of impurity
concentration causes significant change in the persistent current density and
magnetic relaxation.%

\begin{figure}[tbh]%
\centering
\includegraphics[width=8.6cm]{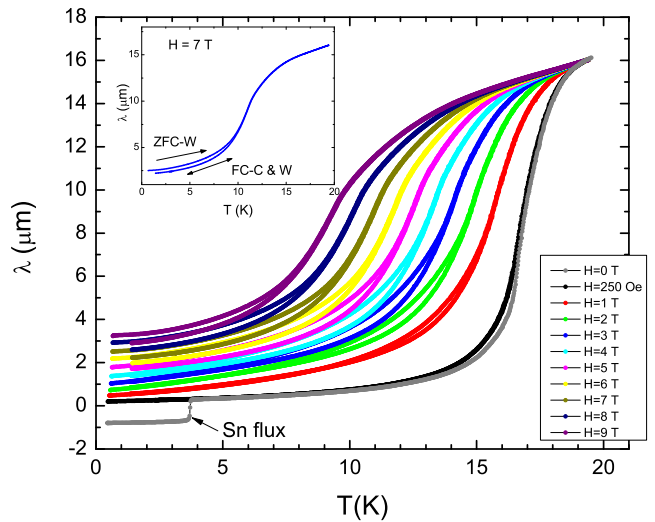}%
\caption{(Color online) Magnetic penetration depth measured in a ZFC-FC process at different
fields. $H=0$ curve shows a step due to leftovers of Sn flux. It was quenched by applying a
$H=250$ Oe field. Inset shows an example of the hysteresis of $\lambda_C(T)$ at $H=7$
T.}%
\label{fig1}%
\end{figure}

In this paper we report measurements of Campbell penetration depth in single
crystals of LiFeAs. We show that the fishtail has dynamic
origin and the field-dependent magnetic relaxation is due to transformation of the pinning potential with field.
Namely, Labusch constant (and \textquotedblleft true\textquotedblright%
\ critical current, $j_{c}\left(  H\right)  $) is a monotonic function of
field when Bean current (macroscopic vortex density gradient) is present, but
it becomes a non-monotonic function of field at a homogeneous distribution of
vortices. The values of
$j_{c}\left(  2\text{ K}\right)  \approx1.8\times10^{6}$ A/cm$^{2}$ provide
upper estimate of the current carrying capability of this material and show
the significance of magnetic relaxation. We also find evidence for the strong
pinning regime at the low fields. With the increase of the magnetic field vortex pinning and creep change to a collective regime and, finally, cross over to another vortex state, perhaps dominated by plastic deformations.
Despite being quite different from high- $T_{c}$ cuprates in terms of pairing and gap structure, it seems
that vortex behavior of Fe-based superconductors is remarkably similar to
high- $T_{c}$ materials.%

\begin{figure}[tbh]%
\centering
\includegraphics[width=8.6cm]{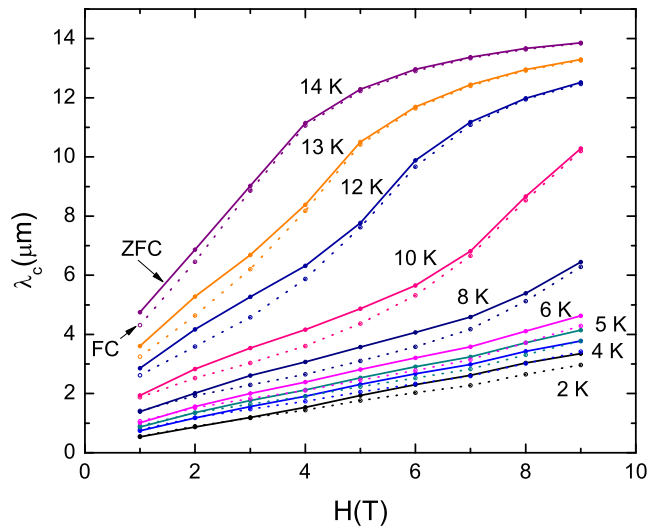}%
\caption{(Color online) Campbell penetration depth as function of magnetic field at different
temperatures extracted from the data of Fig.~\ref{fig1}. Solid lines - ZFC and dashed lines are FC data.}%
\label{fig2}%
\end{figure}

Single crystals of LiFeAs were grown out of Sn flux as described in detail
elsewhere \cite{Lee2010LiFeAsCrystalsSeoul} and were transported for
measurements in sealed ampoules. Immediately after opening, $(0.5-1)\times
(0.5-1)\times(0.1-0.3)$ mm$^{3}$ samples were placed into the cryostat for the
measurements. Additionally, samples were extensively characterized by
transport and magnetization measurements \cite{Lee2010LiFeAsCrystalsSeoul}.
Zero-field transition temperature of our samples was about, $T_{c}\approx
18$~K. The magnetic penetration depth was measured with the tunnel - diode
resonator technique (for review, see \cite{Prozorov2006}). The sample was
inserted into a 2 mm diameter copper coil that produced an rf excitation field
(at $f\approx14$~MHz) of $H_{ac}\sim20$ mOe. An external DC magnetic field
($0-9$ T) was applied parallel to the AC field, both parallel to the $c-$
axis, $H_{ac}\parallel H\parallel c$-axis. The shift of the resonant frequency
(in cgs units) is given by $\Delta f(T)=-G4\pi\chi(T)$, where $\chi(T)$ is the
differential magnetic susceptibility, $G=f_{0}V_{s}/2V_{c}(1-N)$ is a
calibration constant, $N$ is the demagnetization factor, $V_{s}$ is the sample
volume and $V_{c}$ is the coil volume. The constant $G$ was determined from
the full frequency change by physically pulling the sample out of the coil.
With the characteristic sample size, $R$, $4\pi\chi=(\lambda/R)\tanh
(R/\lambda)-1$, from which $\Delta\lambda$ can be obtained
\cite{Prozorov2000c,Prozorov2006}. The measured penetration depth consists of
two terms, London penetration depth and Campbell penetration depth,
$\lambda_{m}^{2}=\lambda_{L}^{2}+\lambda_{C}^{2}$ \cite{Brandt1995}. We
determined $\lambda_{L}\left(  T\right)  $ from the measurements at $H=0$.%

\begin{figure}[tbh]%
\centering
\includegraphics[width=8.6cm]{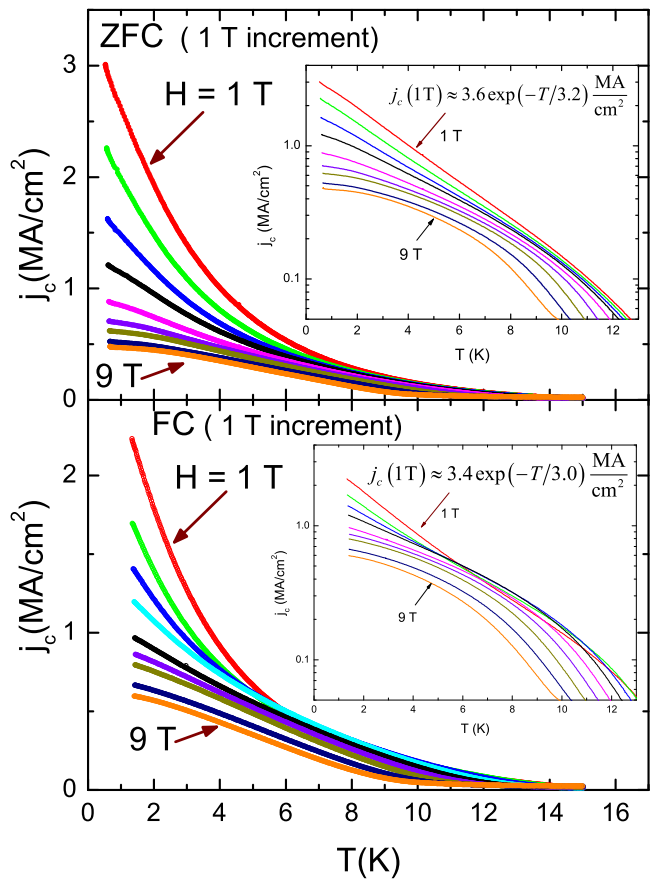}%
\caption{(Color online) \textquotedblleft true\textquotedblright\ critical current density,
$j_{c}$, determined from the ZFC (top frame) and FC (bottom frame) experiments
at indicated values of the applied external magnetic fields. Insets show semi-log plots
indicating exponential dependence of $j_c$ at lower fields and a crossover to a different pinning regime at the higher fields.}%
\label{fig3}%
\end{figure}

Figure \ref{fig1} shows magnetic penetration depth measured upon
warming, after sample was cooled in zero field and target field was applied at
low temperature (ZFC-W) compared to the measurements upon cooling when target
field was fixed above $T_{c}$ and kept constant (FC-C). A step at low
temperatures on a $H=0$ curve is due to residual Sn flux. It was quenched by
applying a moderate $H=250$ Oe field, which does not affect our analysis of
the much higher fields. Inset in Figure \ref{fig1} shows an example of the
magnetic hysteresis measured at $H=7$ T (notice that once ZFC-W process was
complete, subsequent warming-cooling measurements (FC-C and FC-W) resulted in
the same curve indicating homogeneous vortex distribution). The hysteresis
between ZFC-W and FC-C-W is much smaller then, for example, observed in BSCCO
crystals \cite{Prozorov2003}, which is most likely due to much more 3D
electronic nature of LiFeAs. From the measured penetration depth in zero
field, $\lambda_{L}\left(  T\right)  $, and the one measured in applied magnetic
field, $\lambda_{m}\left(  T,H\right)  $, we determine the Campbell
penetration depth via, $\lambda_{C}=\sqrt{\lambda_{m}^{2}-\lambda_{L}^{2}}$ as
shown in Fig.\ref{fig2}.

From the Campbell penetration depth we determine the
\textquotedblleft true\textquotedblright\ critical current density as,
$\frac{4\pi}{c}j_{c}=r_{p}\phi_{0}/\lambda_{C}^{2}$ were we assumed the radius
of the pinning potential be a coherence length, $r_{p}\simeq\xi\simeq7$
nm \cite{Inosov2010}. Figure \ref{fig3} shows $j_{c}$ as a function of temperature at different
magnetic fields determined after ZFC-W process (top frame) and FC-C process
(bottom frame). In both cases, the curves are monotonic in temperature and
show substantial temperature dependence similar to high~$-T_{c}$ cuprates,
re-enforcing the earlier statement that vortex properties of Fe-based superconductors are remarkably similar to the
cuprates, despite the difference in dimensionality of the electronic structure \cite{Tanatar2009b}.

To understand the functional dependence, we plot determined
$j_{c}\left(  T\right)  $ on a semi-logarithmic plot as shown in the insets in
Fig. \ref{fig3}. At relatively low fields, the behavior is very similar to the
earlier reports of strong pinning \cite{Beek2010PrNd1111} and can be well
approximated by the exponential temperature dependence, $j_{c}\left(  1\text{
T}\right)  \simeq3.4\exp\left(  -T/3.0\right)  $ MA/cm$^{2}$ for FC-C process
and $j_{c}\left(  1\text{ T}\right)  \simeq3.6\exp\left(  -T/3.2\right)  $
MA/cm$^{2}$ for ZFC-W measurements. This very similar behavior imply that
strong pins result in a more-or less parabolic $V\left(  r\right)  $ and are
practically independent of the bias Bean current, $j_{B}$. However, at the
higher fields, the critical current becomes less temperature dependent,
probably due to saturation of strong pins and a crossover first to the
collective pinning regime and eventually to the disordered lattice dominated
by plastic deformations.

\begin{figure}[tbh]%
\centering
\includegraphics[width=8.60cm]{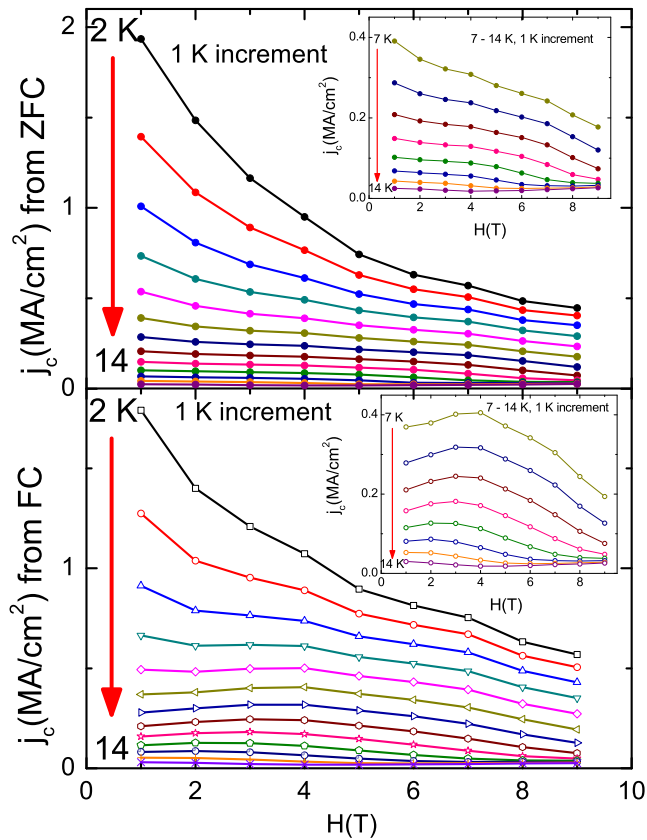}%
\caption{(Color online) Critical current density determined form ZFC Campbell penetration
depth (top frame) and from the FC Campbell penetration depth (bottom frame)
showing the abscence of the fishtail magnetization in the former and its
presence in the latter.}%
\label{fig4}%
\end{figure}

Finally, Fig.\ref{fig4} shows \textquotedblleft true\textquotedblright%
\ critical current density, $j_{c}$, determined form ZFC Campbell penetration
depth (top frame) and from the FC Campbell penetration depth (bottom frame) as
a function of magnetic field at different temperatures. While ZFC curves are
monotonic, a clear fishtail signature is observed in the equilibrium FC-C-W
measurements at higher temperatures. The inset in Fig.\ref{fig4} emphasizes
this result.%

Our results can be interpreted in the following way. Maximum critical
current values, $j_{c}\left(  2\text{ K}\right)  \approx1.8\times10^{6}$
A/cm$^{2}$, show that conventional measurements under-estimate critical
currents, probably due to significant magnetic relaxation. However, the
most striking result is that $j_{c}$, obtained in a non-equilibrium ZFC
process, is monotonic with magnetic field at all temperatures, whereas
equilibrium $j_{c}$ shows a clear signature of the fishtail (second peak)
magnetization. Since conventional (relaxed) DC measurements show fishtail
\cite{Pramanik2010LiFeAs}, we conclude that fishtail effect is of dynamic origin and results from bias (Bean) current - dependent Labusch constant, $\alpha\left(  j_{B}\right)  $. At the critical current density magnetization
is a monotonic function of the magnetic field. However, vortex distribution
relaxes at much faster rates at smaller fields and higher temperatures. During
this relaxation the effective vortex pinning potential transforms, probably
indicating collective effects and ultimately a crossover to the disordered
vortex lattice. In the experiment, it shows as a non-monotonic field dependence of the Labusch
parameter, $\alpha\left(  j_{B}=0,H\right)  $. It is possible that fishtail
has similar static and dynamic origin in high - temperature cuprates and a
very interesting question is how to reconcile very different electronic
properties of Fe-based superconductors and quite similar vortex behavior.

\begin{acknowledgments}
We thank Kees van der Beek, Marcin Konczykowski and Alexey Koshelev for useful
discussion. The work at Ames Laboratory was supported by the U.S. Department
of Energy, Office of Basic Energy Sciences, Division of Materials Sciences and
Engineering under contract No. DE-AC02-07CH11358. Work at SNU was supported by
National Creative Research Initiative (2010-0018300). R.P. acknowledges
support from the Alfred P. Sloan Foundation.
\end{acknowledgments}

\end{document}